# Software supply chain: review of attacks, risk assessment strategies and security controls

Betul Gokkaya[a,*], Leonardo Aniello[a] and Basel Halak[a]

[a] School of Electronics and Computer Science, University of Southampton,
University Road, Southampton SO17 1BJ, UK
*betul.gokkaya@soton.ac.uk, L.Aniello@soton.ac.uk, basel.halak@soton.ac.uk

**Abstract**: The software product is a source of cyber-attacks that target organizations by using their software supply chain as a distribution vector. As the reliance of software projects on open-source or proprietary modules is increasing drastically, SSC is becoming more and more critical and, therefore, has attracted the interest of cyber attackers. While existing studies primarily focus on software supply chain attacks' prevention and detection methods, there is a need for a broad overview of attacks and comprehensive risk assessment for software supply chain security. This study conducts a systematic literature review to fill this gap. We analyze the most common software supply chain attacks by providing the latest trend of analyzed attacks, and we identify the security risks for open-source and third-party software supply chains. Furthermore, this study introduces unique security controls to mitigate analyzed cyber-attacks and risks by linking them with real-life security incidence and attacks.

**Keywords:** Software Supply Chain, Security, Risk, Attack, Security Controls

## 1. Introduction

Software Supply Chain (SSC) refers to the set of processes to select and obtain software components from third parties; it also encompasses the companies involved in these processes. As the reliance of software projects on open-source or proprietary modules is increasing drastically, SSC is becoming more and more critical and, therefore, has attracted the interest of cyber attackers. Indeed, Gartner predicted that "by 2025, 45% of organizations worldwide will have experienced attacks on their software supply chains"[1]. Recently, cyber-attacks have been targeting companies by infiltrating their SSC, for example, to inject malicious code into the software used. The SolarWinds attack [1] is a noteworthy example of this kind, where adversaries weaponized a network monitoring software developed by SolarWinds and used by several thousands of customers, including large companies and government entities; as a result, the attackers gained remote access to the IT networks of the affected organizations.

The software used by enterprises, governments, and national critical infrastructures includes dependencies on modules developed and maintained by many different suppliers. Therefore, each software supplier in the SSC is a potential entry point for attackers to target key national assets in cyberspace. With reference to the National Cyber Strategy 2022[2], cyber resilience depends on the ability to (i) understand the nature of the risk, (ii) take action to protect systems and withstand cyber-attacks, and (iii) minimize the impact of successful cyber-attacks. With such a comprehensive, complex, and growing SSC attack surface, an accurate assessment of risk is extremely challenging to perform. This significantly hinders the ability to understand SSC risk and, consequently, secure systems and resist attacks.

[1] https://www.gartner.com/en/articles/7-top-trends-in-cybersecurity-for-2022

[2] https://assets.publishing.service.gov.uk/government/uploads/system/uploads/attachment_data/file/1053023/national-cyber-strategy-amend.pdf - paragraph 91

Despite the growing interest from academia in addressing security challenges within SSCs, particularly in the open-source domain due to its public accessibility, comprehensive risk assessment for both open-source and third-party supply chains remains a crucial-yet under-explored area. Current research predominantly centres on open-source software supply chain attacks, likely because the open-access nature of these projects enables researchers to assess the security threats or attacks targeting them more easily. However, a broader understanding of risks in various software supply chain scenarios encompassing open-source and third-party contexts is essential for more effective security risk management.

Several studies have evaluated the security risks associated with software supply chains, focusing on open-source development environments or package managers, such as npm. However, the number of risk assessment studies is relatively limited compared to those investigating security attacks against SSCs. Zahan et al. investigated 1.63 million JavaScript npm packages for the open-source software chain and proposed six vulnerable entry points that an attacker can use to exploit SSC security [2]. However, their security analysis is limited to addressing widened SSC challenges and only proposes vulnerable entry points for open-source software chains, specifically for JavaScript. Wang et al., studied the third-party libraries' usage, updates, and risks for Java projects [3]. The study provides library usage analysis for open-source and third-party libraries to carry on a library risk analysis. Although the work benefits researchers in understanding the risks of using third-party libraries, it fails to address extensive SSC challenges and risks. Zimmermann et al., focus on security risks in the context of the npm package manager, revealing a considerable attack surface due to the densely connected nature of the ecosystem. To address these risks, the researchers propose mitigation strategies, such as vetting trusted maintainers and assessing new package releases' code. Ladisa et al., present a valuable contribution to the field by providing a comprehensive taxonomy of attacks on OSS supply chains, validated by domain experts.

While these papers make valuable contributions to understanding OSS supply chain attacks, a comprehensive assessment or overview of available techniques for addressing supply chain risk assessment for both open-source and third party SSCs is currently lacking in the literature. Our study encompasses the risk assessment and offering an overview of the available techniques for this purpose. By leveraging a systematic analysis of academic literature, grey literature, and an open-access dataset, we offer a more holistic understanding of the risks involved in software supply chain management. This contribution enables organizations to better assess and mitigate potential risks across a broader spectrum of supply chain scenarios, thereby strengthening their overall security posture in the face of an evolving threat landscape.

More specifically, the contributions of this research are as follows:

- We conducted a systematic literature review to analyse various attack vectors and identify common patterns in SSC attacks. In addition, we utilized the Atlantic Council's SSC attacks dataset, comprising 161 recorded incidents from 2010 to 2021, to determine the latest trends in SSC attack strategies.
- We synthesized diverse risk assessment approaches and examined their applicability to software vendor activities. This comprehensive risk assessment framework offers valuable insights into the complexities of managing software supply chain risks.
- Based on the literature analysis and the proposed risk assessment approaches, we provide a set of practical recommendations for implementing security controls. These controls, when adopted by organizations, can mitigate the likelihood of SSC attacks and enhance overall software supply chain security.

The rest of this report is organized as follows. Section 2 provides an introduction to software supply chains. Section 3 illustrates the methodology used in this study for literature analysis. Section 4 breaks down and explains the most common types of supply chain cyber-attacks. Section 5 gives an overview of the latest trend in SSC attacks. Section 6 presents a review of SSC risk assessment approaches. Section 7 derives a set of control measures that organizations can implement to address SSC risk. Finally, the conclusion and suggestions for future work are provided in Section 8.

# 2. Software Supply Chains

Most companies use software applications developed by vendors or released as open-source software (OSS). Each software application consists of several components, some developed by the software vendor or OSS team directly, others supplied by third parties, and imported as dependencies. Also, imported components can depend on other components provided by additional suppliers, and the chain of dependencies may extend to include further suppliers (see Figure 1) [4].

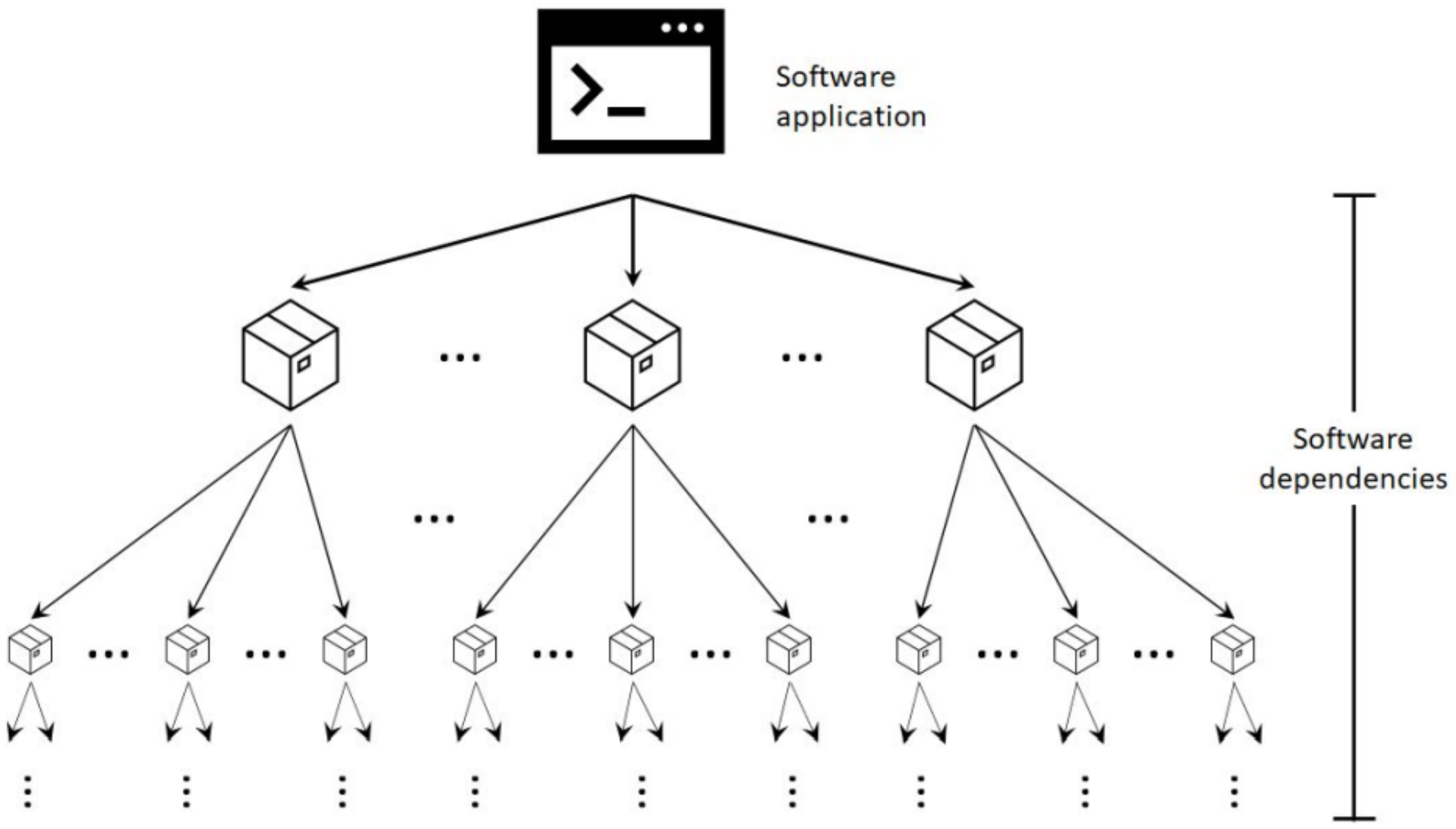

*Figure 1 - an illustration of a chain of dependencies for a single software application*

Although different software development approaches exist, most of them include a *development stage*, a *build stage*, and a *distribution stage*, as shown in Figure 2. In the development stage, developers write the source code and commit updates (e.g., a software security update) to the source code management system used by the company. Emphasizing the importance of security and best practices during the development stage is necessary to minimize vulnerabilities and prevent potential attacks. By incorporating security measures, such as input validation or secure coding techniques, developers can reduce the risk of code injection attacks, exploitation of unmaintained legacy code, and other security threats, which will be explained in Section 4.

In the build stage, maintainers are in charge of operating the build management system which combines the code developed in-house with other components developed by third parties. These components are imported as packages and can be provided by either other software development companies or by OSS projects. For each software project, the list of dependencies used is usually managed by a package manager, included within the build management system. Different programming languages use different package managers to handle and import dependencies, e.g., npm[3] for JavaScript, and Maven[4] for Java. At the end of the build stage, the software product is ready to be distributed to users and

[3] npm https://www.npmjs.com/

[4] Maven https://maven.apache.org/

customers, via some distribution means such as a software repository (e.g., Google Play) or website.The SSC for a software application encompasses all the suppliers that contributed to at least one of the components of that application (see Figure 2). The SSC of a company is the union of the SSCs of all the software applications in use within the company's software estate.

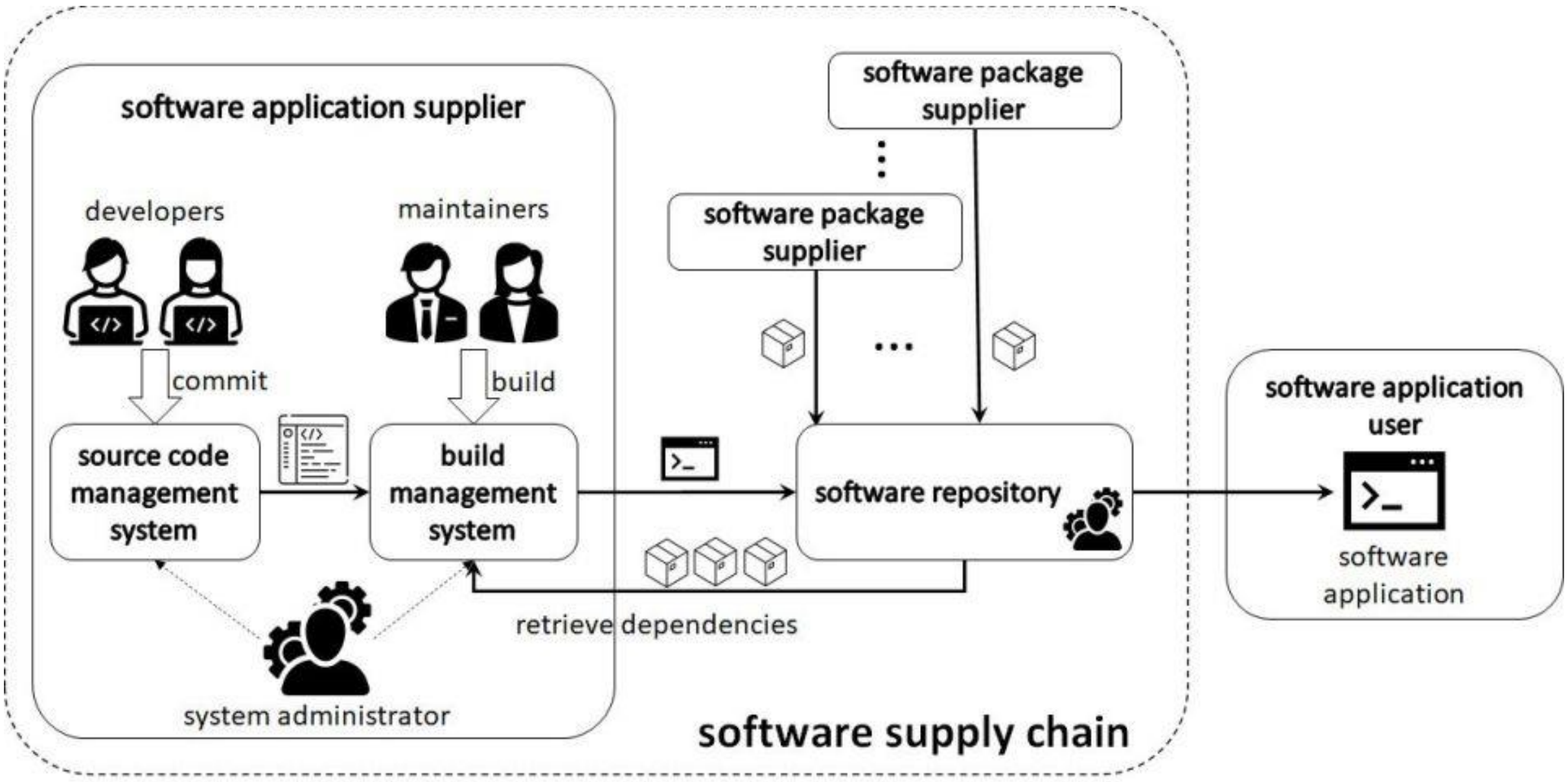

*Figure 2 - software supply chain (SSC) of a software application*

A company may also rely on remote services provided by cloud providers, such as email (i.e., Software as a Service, SaaS) or virtual machine renting (i.e., Infrastructure as a Service, IaaS). In this case, the actual software application providing the cloud services is hosted by the provider and consumed by the company remotely. In this report, any cloud service used by a company is considered part of its SSC. Hereinafter, we refer to software application vendors and cloud providers as *software providers*, and to companies ingesting software or using digital services as *software users* (see Figure 3).

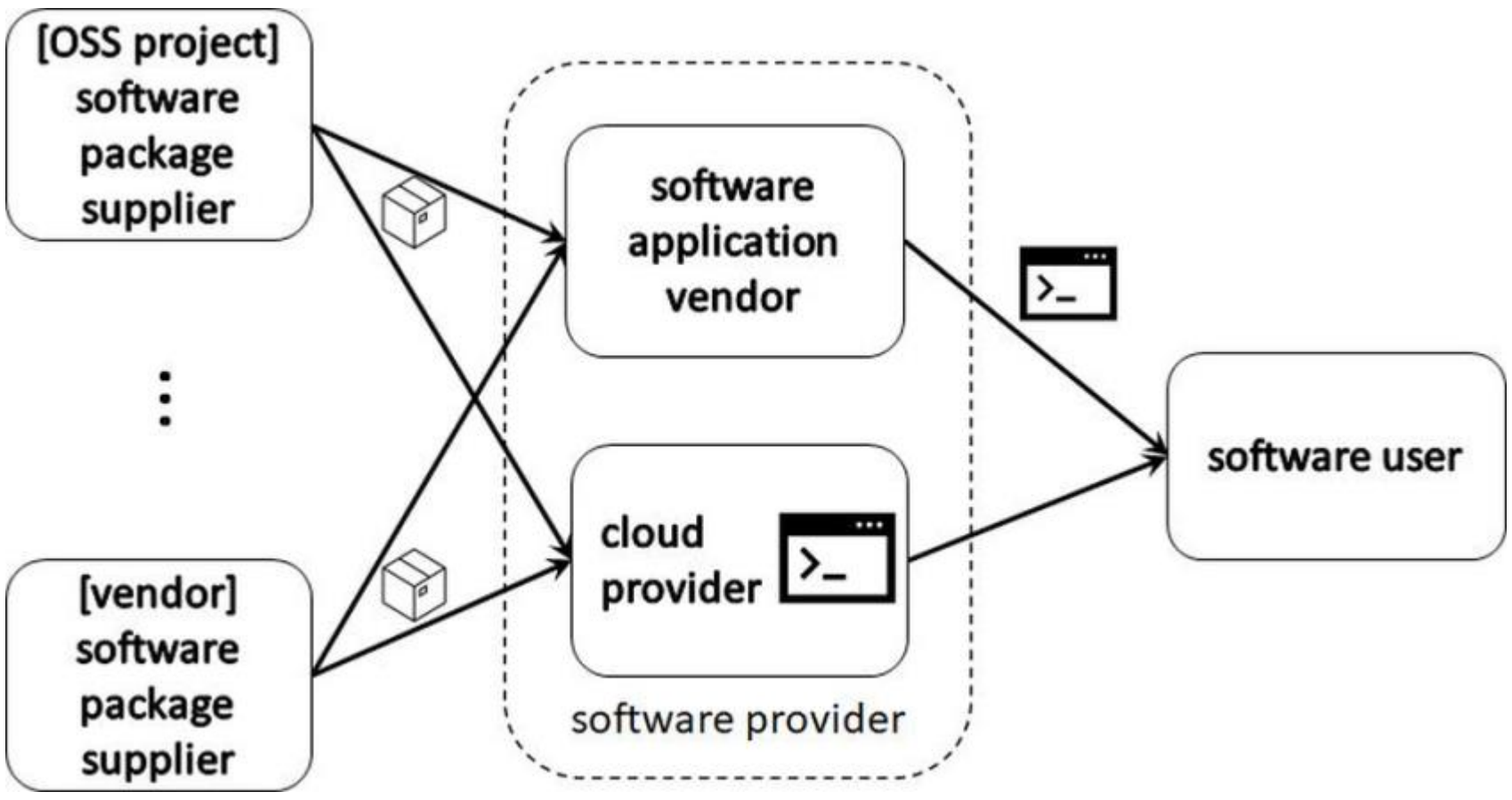

*Figure 3 – relationships between a software user (either ingesting software or using a cloud service) and its software provider (either a software application vendor or cloud provider), as well as the software package suppliers of the latter, which can be either open-source projects or other software, vendors.*

# 3. Literature Review Methodology

This section explains the literature review methodology adopted in this study. The methodology includes two steps: (1) identifying the research scope and selecting the research articles and grey literature, and (2) analyzing the selected literature. The first step is *identifying the research scope and selecting the research articles and grey literature*.

This study focuses on peer-reviewed research articles published in the English language between 2015 and 2022. We limited our literature review to peer-reviewed articles to ensure the academic quality of selected studies. Moreover, the studies published before 2015 were omitted because they were not considered representative of the current SSC threat landscape. We used the identified search terms "software supply chain attack", "software supply chain security risk", and "software supply chain security" in the following databases, which are recognized sources of academic literature: Google Scholar (205 results), ACM Digital Library (75 results), and IEEExplore (10 results). After that, we removed duplicated papers, which resulted in 222 papers left from the original 290. Then, we applied the following inclusion criteria.

1. Journals with an impact factor of at least 2, and conferences rank at least B
2. Research limited to journal and conferences article for scientific literature
3. Research limited to whitepapers, blog posts, and annual security reports for grey literature

In this study, the journal papers with an impact factor of at least 2 and conference papers with a rank of at least *B* were chosen to ensure that selected studies are of appropriate quality. After reading the abstracts of all selected papers, we removed 22 articles because they were not directly related to SSC attacks or risks. Afterward, we employed the *snowballing technique* to also include those articles that were cited by the selected papers. We had 71 papers after applying the *snowballing technique*. Finally, we applied again the inclusion criteria mentioned earlier and reduced the set of articles to 53.

# 4. Cyber Attacks via Software Supply Chains

SSC attacks aim at compromising the cyber infrastructure of a company by targeting software applications, or their delivery mechanisms before the company itself receives them. In this sense, the attacker exploits the SSC to deliver a malicious or vulnerable payload to a target organization. Once this payload has been deployed and activated at the target's premises, the attacker can move on to the next stages of the attack, such as exploitation, command and control, lateral movement, and data exfiltration [5]. This section focuses solely on the attack tactics and techniques used by malicious actors to infiltrate an SSC to deliver a malicious or vulnerable software component.

A company may use a software application that has a vulnerability introduced due to bad software development practices by a supplier. In this case, an attacker does not need to target the SSC and can directly target the company by exploiting that vulnerability [6]. This does not classify as an SSC attack, because the attack would take place after the software application is delivered to the company. However, since the vulnerable software application represents a threat to the cyber security of the company and is delivered through its SSC, this kind of threat is taken into account for SSC risk assessment (see Section 5).

## 4.1. Typosquatting and combosquatting attacks

Typosquatting refers to the practice of producing and registering a malicious package with a name that is a misspelling of a well-known legitimate package, which may mislead users to download the malicious one from package managers into their application [7]. The open-access feature of package managers like npm allows anyone to upload packages with any chosen name [8]. Usually, the mistyping consists of deleting or inserting hyphens, or excluding or switching letters [9]. Table 1 reports some real examples of typosquatting attacks.

When selecting and importing open-source dependencies into a software project, developers or maintainers may fall victim to this attack and include malicious packages in their software code [10]. Victims of typosquatting attacks can remain unaware of having included a malicious dependency for months or even years [7]. They can build new dependencies upon the malicious ones, resulting in widened attack surfaces. Taylor et al. [7] showed that a malicious version of the popular *lodash* package was published with the name *loadsh*. The authors contacted the potential targets of this attack, and three *loadsh*-dependent package managers reported that they installed the malicious package without realizing it. Furthermore, as well as package managers downloading the malicious package directly, several downstream developers were also affected as they ingested the loadsh package as a dependency of another package.

Previous studies indicate that the typosquatting technique is simple to implement and common in practice [11], [12]. Ohm et al., [9] found that typosquatting has been used in more than half of the 174 malicious packages (61%) used in their selected corpus of SSC attacks. Consequently, statistics and real-life incidents exemplify the effectiveness of the typosquatting technique.

Combosquatting is another version of a name similarity attack where a threat actor may try to deceive the user with a similar name to the original package of the malicious version [13], [14]. In this attack technique, malicious packages can be distributed by manipulating the common packages' names by adding common prefixes or suffixes, e.g., *pytz* into *pytz2-dev*, creating a malicious version of the original package which, in this example, is the distribution of a Python 2 development version [11]. Moreover, the singular version of the legitimate package name is another way of tricking the users into installing the original plural form of the package, such as *master tool* instead of *master tools*. For more examples, Table 1 illustrates past combosquatting and typosquatting attacks in PyPI.

| Year | Malicious Package | Original Package | Changes |
|---|---|---|---|
| 2019 | python3-dateutil | python-dateutil | add ‘3’ |
| 2019 | jellyfish | jellyfish | Substitute ‘l’ with ‘I’ |
| 2018 | pythonkafka | kafka-python | Swap “kafka” and “python” & remove ‘-’ |
| 2018 | python-mysqldb | MySQL-python | Swap “python” and “MySQL” & add “db” |
| 2018 | libhtml5 | html5lib | Swap “html5” and “lib” |

*Table 1 - Combosquatting and typosquatting attack examples in PyPI (adapted from* [11]*)*

## 4.2. Code Injection Attack

Code injection occurs when the threat actor injects malicious code into a piece of software to exploit it at a later stage [9], [15], [16]. Sometimes, code injection attacks involve changing only a tiny portion of the legitimate artifact, which can be highly challenging for developers to detect. In this way, the attackers can stealthily "set traps" upstream, allowing them to carry out attacks against the end-users further down the chain [13], [17].

Attackers can inject malicious code into a project during the source code development process by obtaining maintainers' credentials. For example, the *rest-client*[5] attack was carried out by accessing a maintainer's account by exploiting weak credentials (i.e., an insecure password was used). The attackers installed a backdoor and more than ten libraries were compromised before the malicious code was detected and removed [18]. Identifying malicious code in the project code can be challenging due to its silent nature until the attacker wants to exploit them [19], [20], [21].

## 4.3. Exploiting Unmaintained Legacy Code

Popular package managers (e.g. npm) may include libraries that have numerous unfixed vulnerabilities [22]. For example, in February 2022 it was reported over the past six months, 1,300 malicious packages were detected in the npm package manager[6]. Some of those vulnerabilities can remain unfixed even after they have been detected because either the package is no longer maintained, or the corresponding patches have not been released [23]. Thus, an attacker can take advantage of vulnerable unmaintained codes to compromise the applications that import them as dependencies [8]. This attack usually takes place once ingested by the software user (see Figure 2) and does not target directly any previous stage of the SSC.

## 4.4. Ownership Transfer

The ownership transfer technique refers to the attacker convincing a legitimate package maintainer to register them as an additional maintainer. Once the attacker becomes a legitimate maintainer, they can take over the targeted package by removing the other existing owner [8]. The ownership transfer technique is also referred to as package takeover in the literature [8]. Once a threat actor becomes the only maintainer of a project, they can easily inject malicious code which would then be disseminated to all the software projects that use it as a dependency[24].

The *event-stream*[7] incident is a relevant real-world example of this attack, where the adversary was granted publishing rights on the event-stream package by using social engineering techniques against legitimate maintainers [25] . The attacker then removed all the original maintainers of the event-stream package, injected malicious code, and published it to the package repository [26].

[5] https://www.securityweek.com/backdoor-found-rest-client-ruby-gem
[6] https://www.securityweek.com/1300-malicious-packages-found-popular-npm-javascript-package-manager
[7] https://github.com/dominictarr/event-stream/issues/116

## 4.5. Account Takeover

Attack takeover consists of an attacker obtaining access to the account of a developer or maintainer of the targeted packages [2],[8]. Unlike the ownership transfer attack, with this technique, the malicious actor targets the developer or maintainer account directly, instead of aiming to become another package maintainer. Social engineering techniques are used to take over the account. This attack can be performed using different techniques, e.g., by exploiting weak or compromised passwords or social engineering. For example, in 2019, the account of a developer of the popular *rest-client* package was compromised to install crypto miners, and infected versions were installed about one thousand times[8].

## 4.6. Metadata manipulation attacks

The metadata manipulation attack was introduced by Torres-Arias et al. [27] as a new threat class against Git to manipulate its metadata stored in the *.git/refs* directory. Git is open-source software that is free for developers to use for coordinating work and open-source project revision[9]. This attack aims to trick different developers who have different views about the state of the same software repository, which leads them to take actions that can undermine the security of the software itself. For example, developers may omit security updates, merge untested code, or import vulnerable code [28]. To perform this attack, the attacker takes advantage of the lack of a mechanism in Git to protect its metadata, e.g., not requiring metadata to be cryptographically signed [27].

## 4.7. Certificate theft

Software developers sign the code updates they commit using digital certificates, to assure that the code is legitimate and trusted. Attackers can steal these certificates to sign malicious updates and mislead users into importing infected or vulnerable packages [29].

## 4.8. Exploitation of signature system

An attacker can achieve the same effects of stealing a legitimate certificate (see Section 4.8) by exploiting the signature mechanisms used within the software development pipeline. Different techniques can be employed for this purpose. For example, some OSS repositories do not enforce proper signature verification, therefore an attacker can simply push malicious updates without resorting to any attack technique. As another example, the signature system itself might be broken and skip the signature verification step. Finally, the adversary might breach a software provider and inject malicious code before the software is signed.

# 5. Latest trends in SSC attacks

An analysis of recent SSC attacks helps highlight the approaches that attackers have been using most frequently, from which we can identify which techniques are most likely to be employed. Most data on SSC attacks are found in the grey literature and a common theme is a prominent focus on open-source repositories used as the main distribution vector. This is partly due to the wider availability of data related to open-source projects compared to proprietary software. This data shows that the number of SSC attacks that exploited weaknesses in open-source projects grew by 430% in 2020 and by 650% in 2021 [18]. Also, more than one report reveals that most software nowadays, including those developed by third-party companies, include many open-source components [17][30]. The staggering increase of SSC attacks having OSS projects as entry points, together with the observation that nowadays software is largely made by OSS components, suggests that addressing risk from OSS projects is a priority.

To extract further statistics on the latest SSC attacks at a global level and identify the most used approaches, we relied on the SSC attacks dataset provided by the Atlantic Council, which includes 161 attacks from 2010 to 2021 against companies worldwide. To identify the latest trends, we only selected attacks from 2018 onwards (106 in total). As shown in Figure 4, the top three distribution vectors cover the majority of SSC attacks: open-source dependencies (23%), supply chain service providers (20%), and hijacked updates (15%). While the first one aligns with the finding we highlighted in the previous paragraph, the other two suggest that software providers are commonly used as entry points for SSC attacks.

[8] https://www.helpnetsecurity.com/2019/08/21/backdoored-ruby-gems/
[9] https://git-scm.com/book/en/v2/Getting-Started-What-is-Git%3F

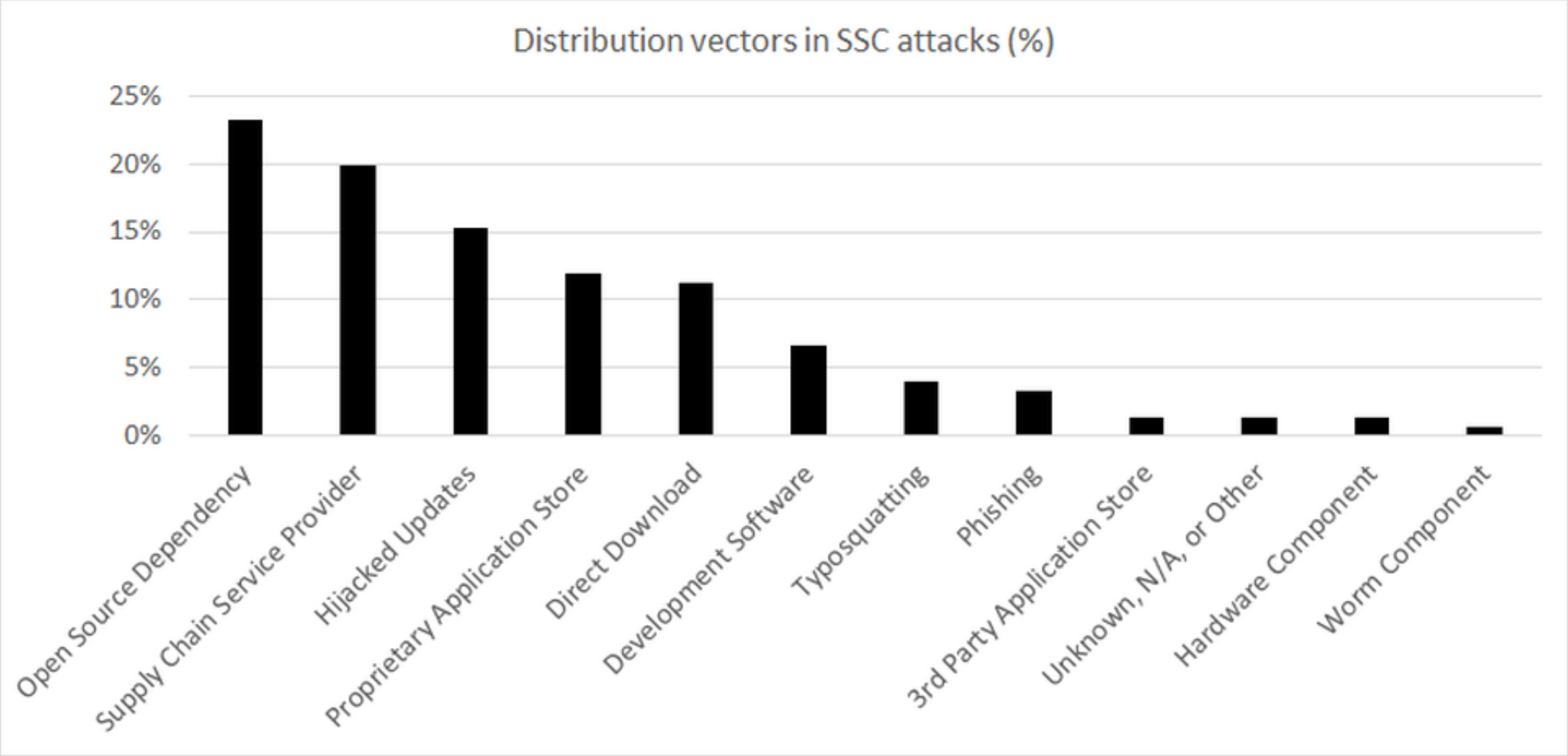

*Figure 4 – Frequency of distribution vectors in SSC attacks in 2018-2021*

In terms of attack vectors, Figure 5 shows that the top three known vectors (i.e., excluding the "Unknown, Other, or N/A" category) represent the majority of considered attacks: exploitation of the lack of proper signature verification of OSS packages (34%), exploitation of vulnerabilities to bypass authorization or signature verification (12%) and unauthorized access to accounts that have the permissions to inject code in the pipeline, usually by exploiting cyber or human vulnerabilities in software vendors or cloud service suppliers (11%).

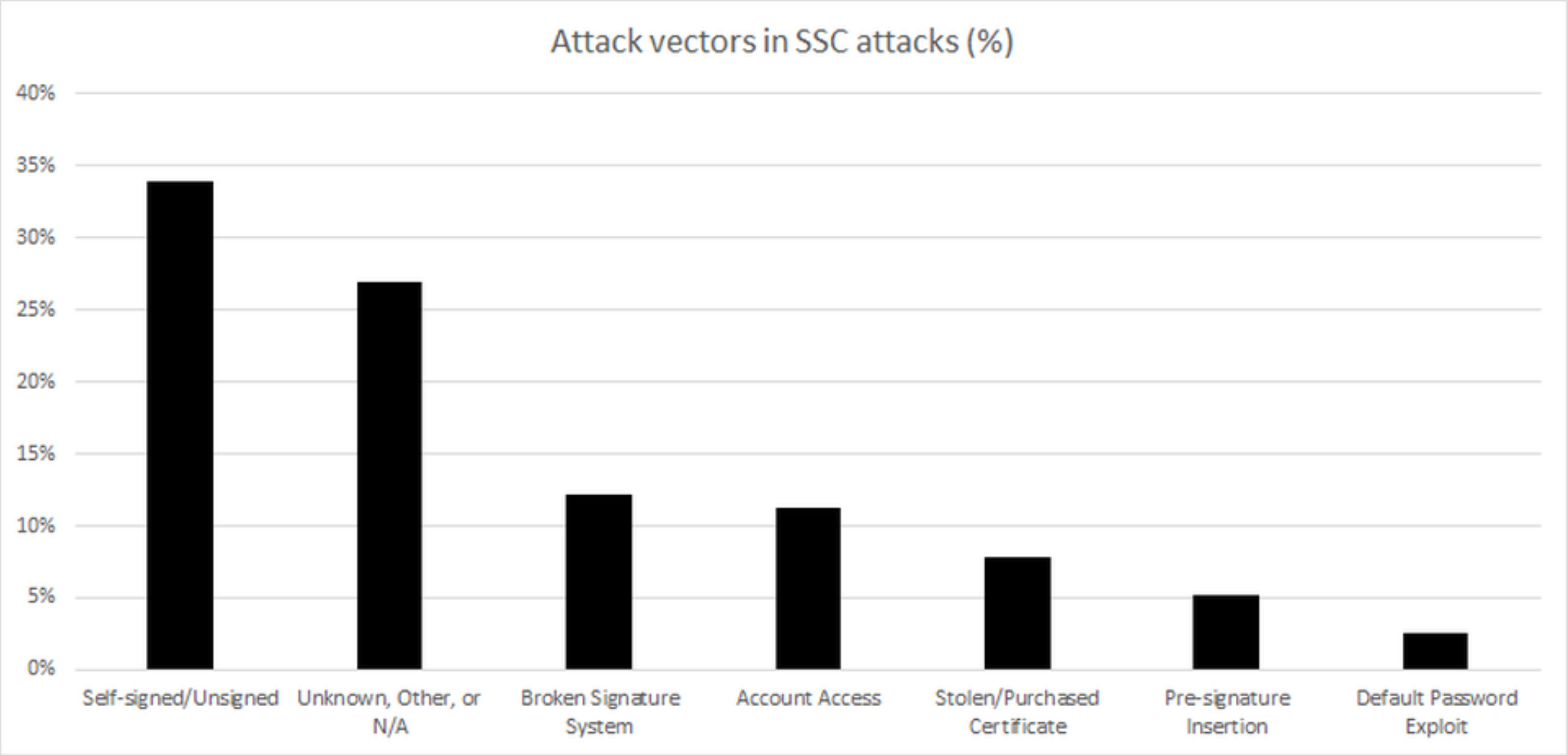

*Figure 5 – Frequency of attack vectors in SSC attacks in 2018-2021*

To sum up, the most likely SSC attack techniques to be considered for risk management are those which use as entry point either (i) OSS projects, (ii) software vendors, or (iii) cloud service suppliers, and those which use as attack technique the exploitation of either (i) the lack of proper signature verification of OSS packages or (ii) vulnerabilities to bypass authorization or signature verification or (iii) cyber/human vulnerabilities in software vendors or cloud service suppliers.

# 6. Software Supply Chain Risk Assessment

As discussed in Section 4, the attack surface of the SSC is extensive and offers attackers several possible entry points to target organizations. A risk assessment strategy is therefore required as a first step to address the many threats deriving from the SSC. The end goal of an SSC risk assessment is to identify these threats and prioritize them based on (i) the perceived likelihood they would occur and (ii) the estimated impact they would cause in case they occurred. The benefits of an SSC risk assessment include an increased awareness of potential harm from used software applications and dependencies, as well as an indication of which threats should be addressed more urgently.

Although none of the reviewed papers proposes a comprehensive strategy for SSC risk assessment, some articles introduce relevant metrics and approaches that can be used to estimate specific aspects of SSC risk. Most of these papers fall into one of the following categories: they either focus on the assessment of vulnerabilities in used software and imported dependencies or the security evaluation of an OSS project [31]. The primary focus of the first group's studies is to identify the vulnerabilities in reused software, dependencies, or package managers by implementing various methods, such as static analysis, code review, and implementing security metrics (more information provided in Table 2 column 6) to accomplish this goal. Among the reviewed studies, Java, JavaScript, and Python are the main investigated programming languages, while the most popular package managers may include Maven, npm, RubyGems, and PyPI.

On the other hand, the second group studies propose new methods or tools to improve the integrity of the open-source software supply chain. The studies in this category are dedicated to analysing, developing, and implementing tools, methodologies, and techniques for assessing and enhancing the security of open-source software projects. By addressing various aspects of software security and supply chain management, these studies contribute to a more robust and secure OSS ecosystem. Further information about the objectives and technical implementation of the reviewed studies is shown in Table 2.

| Reference | Primary Focus | Objective | Method | Implemented analysis | Language /Package manager | Outcome |
|---|---|---|---|---|---|---|
| [31] (2021) | Vulnerability assessment in used software | The study aims to understand the association between security vulnerabilities and code reuse in open-source software systems. | The study analyses a dataset of 1244 Java projects. | Static analysis | Java | 65% of the analysed projects have at least one security vulnerability introduced through a dependency. |
| [32] (2018) | Security evaluation of an OSS | The study aims to develop a tool that improves security in products for better manage vulnerabilities and patching of IoT devices and standalone applications. | The tool matches software components and release versions with vulnerability information from open resources. | none | General | The tool helps users and organizations to understand the current security status of their devices and work more efficiently with vulnerabilities. |
| [33] (2018) | Vulnerability assessment in imported dependencies | This study aims to develop a precise approach for measuring vulnerable dependencies in open-source software. | The study analyses 10,905 library instances of the 200 Java Maven-based open-source libraries most frequently used in SAP software. | Dependency analysis | Java | The results showed that 20% of dependencies with known vulnerabilities were not actually deployed. Plus, developers were able to directly address 82% of their vulnerable dependencies, representing a 45% improvement in comparison to conventional methods. |
| [10] (2020) | Vulnerability assessment in imported dependencies | This study aims to explore the issue of typosquatting and combosquatting attacks in PyPI. | The study employs the *Levenshtein distance*[10] to identify packages that are potentially involved in combosquatting and typosquatting attacks. | Manual analysis | Phyton | The study found 62 packages with the same name as any of the 297 module names in the Python standard library, which were mostly kept in PyPI for backporting reasons. |
| [19] (2022) | Vulnerability assessment in used software | This study focuses on how the security releases of open-source packages can result in attacks. | The study analyses a dataset of 4,377 security advisories across seven package ecosystems. | Code analysis | Composer, Go, Maven, npm, NuGet, pip, RubyGems | The median security release becomes available within four days of the corresponding fix and contains a 131 lines of code change. However, one-fourth of the releases take at least 20 days after the fix. |
| [34] (2022) | Security evaluation of an OSS | The study introduces a novel attack scenario where an SGX binary enclave is compromised before the enclave undergoes measurement, and the subsequent measurement is incorporated into the signing material. | The study proposes a new attack scenario that corrupts an SGX binary enclave before the enclave is measured and the resulting measurement is included in the signing material. | Security testing | None | The analysis reveals that SGX protection measures can be practically nullified by the proposed malware attack. |
| [16] (2020) | Security evaluation of an OSS | This study aims to propose a lightweight and efficient approach for detecting code injections in software packages. | The study offers a new method to identify the code injection attack. | Static analysis | Python | The analysis of 2,666 software artifacts from the top ten most downloaded Python packages in PyPI shows that the proposed technique is applicable for lightweight analysis of real-world packages. |

[10] https://www.cuelogic.com/blog/the-levenshtein-algorithm

| | | | | | | |
|---|---|---|---|---|---|---|
| [35] (2022) | Vulnerability assessment in used software | This study introduces the concept of "shrinkwrapped clones" in the npm software package ecosystem and presents a mechanism called "unwrapper" to detect these clones and match them to their source package. | The study uses a package difference metric based on directory tree similarity, along with a prefilter to eliminate packages unlikely to be clones of a target. | Static analysis | Npm package manager | The research scrutinizes the comprehensive npm ecosystem, encompassing 1,716,061 packages and 20,190,452 distinct versions, ultimately unveiling 6,292 heretofore undiscovered shrinkwrapped clones. |
| [36] (2020) | Security evaluation of an OSS | The study addresses the challenge of locating affected API uses in client code when adapting to non-backwards compatible changes in JavaScript libraries. | The study presents a pattern language for describing the API access points involved in breaking changes and an accompanying program analysis tool called TAPIR. | Static analysis | JavaScrip t | The results show that the pattern language is expressive, with 187 breaking changes from 15 package updates expressed using a total of 283 patterns. |
| [22] (2022) | Vulnerability assessment in used software | The study investigates the spread of vulnerabilities through code reuse and the potential impact of these secondary vulnerabilities. | VDiOS is developed to identify and fix white-box-reuse-induced vulnerabilities previously resolved in the originating projects (termed orphan vulnerabilities). | Case study | GitHub, Bitbucket, Source-Forge | The study finds multiple orphan vulnerabilities in active, widely acclaimed projects (exceeding 1,000 stars) and inactive projects. The frequently long duration required to address these orphan vulnerabilities in highly popular endeavours amplifies the probability of their propagation to emerging projects. |
| [37] (2022) | Security patch identification | The study offers a security patch identification system, termed E-SPI, which encapsulates the structural data inherent in a commit for proficient identification purposes. | The new method, E-SPI, retrieves the hidden structural data embedded within a commit to ensure sufficient identification. | none | General | The new method shows a 4.01% enhancement in accuracy and a 4.22% improvement in the F1 score utilizing the prevailing dataset. |
| [7] (2020) | Vulnerability assessment in used software | The study develops a tool called *TypoGard* for detecting and reporting potential typosquatting incidents in software repositories before they can impact users. | TypoGard identifies unpopular packages with lexically similar names to popular ones, leveraging a model of lexical similarity and incorporating package popularity. | Security testing | npm, PyPI and RubyGe ms | The shows that TypoGard detects up to 99.4% of known typosquatting cases while generating a small percentage of warnings up to 0.5% of package installs and low overhead package install time (2.5%). |
| [38] (2022) | Security evaluation of an OSS | The study develops an OSS Abandonment Risk Assessment (OSSARA) model for identifying components integrated into software products with a high probability of imminent abandonment. | OSSARA computes the abandonment risk for a software system by considering 1) the probability of each OSS component losing maintenance support within a specified timeframe, and 2) the significance of each component in relation to the primary system. | Security testing | General | OSSARA offers a monitoring system for organisations in order to identify risk levels and choose to maintain or replace OSS components as needed. |
| [13] (2022) | Vulnerability assessment in used software | The research presents a technique for identifying code injections in npm packages by examining inconsistencies between source code and the corresponding package. | A dataset ([9]) comprising malicious npm packages serves as the basis for evaluating the proposed method, and its efficacy is contrasted with the alternative git-log solution to determine relative performance. | Vulnerability detection | Npm-JavaScript | The suggested implementation demonstrates efficiency (20.7 times swifter than git-log) and scalability, accommodating sizable packages such as the commander package. |
| [39](2018) | Security evaluation of an OSS | The study introduces a methodology for the automated detection of security-related alterations in source code. | The research regards source code modifications as documents composed in natural language, subsequently categorizing them utilizing conventional document classification techniques. | Machine learning-based document classification | Java project used in SAP software | The recommended technique achieves 80% precision and 43% recall, providing a substantially reduced volume of training data and employing a more straightforward architecture in contrast to prevailing approaches. |
| [40] (2016)<br>[40] (2016) | Security evaluation of an OSS | This study presents Diplomat, a practical security system for community repositories that provides immediate project registration and compromises resilience. | The research uses *prioritized delegations* to impose a hierarchy among parties with equal trustworthiness and utilizes terminating delegations to inhibit statements originating from less trusted entities from being deemed trustworthy for a given package. | Delegation techniques, prioritized and terminating delegations | Ruby, CoreOS, Haskell, OCaml, Python, Flynn, LEAP, and Docker. | The study demonstrates that Diplomat can safeguard more than 99% of PyPI's users, even when an attacker obtains control over the repository for one month. |
| [8] (2019) | Vulnerability assessment in imported dependencies | This study analyses security risks npm users face by systematically exploring dependencies among packages, the maintainers accountable for these packages, and publicly disclosed security issues. | The study analysed the security threats that pose a high risk in npm and identified those threats by investigating package dependencies, and maintainers and recognised security vulnerabilities within npm. | Security testing | npm | On average, installing npm package entails implicit trust in 79 third-party packages and 39 maintainers, thereby establishing an unexpectedly vast attack surface. Highly popular packages, either directly or indirectly, impact numerous other packages (frequently surpassing 100,000) The outcomes show that npm is afflicted by single points of failure, and that neglected packages present risks to extensive code bases. |
| [41] (2021) | Vulnerability assessment in imported dependencies | This study analyses the categories of alterations potentially impacting open-source dependencies and evaluates their effect on the efficacy of vulnerability detection tools. | The study encompasses the examination of 7,024 Java projects using Eclipse Steady, OWASP Dependency Check, and a commercial vulnerability scanner and, it categorizes vulnerabilities identified for the top 20 most-utilized dependencies into true and false positives. | Case study | Java project used in SAP software | The work proposed that the identified modifications cause a significant obstacle in identifying vulnerable open-source software, which also shows that none of the examined scanners can efficiently handle all modification types. The study underscores the necessity for additional research to enhance detection algorithms and techniques employed by vulnerability scanning tools. |
| [42] (2022) | Vulnerability assessment in imported dependencies | The study represents the method called Vuln4Real to count vulnerable dependencies in Maven. | The research examines the top 500 prevalent open-source Java libraries employed by SAP within its proprietary software, with the evaluation encompassing 25,767 distinct library occurrences in the Maven repository. | Dependency analysis | Java libraries used in SAP software | The study illustrates that the proposed method substantially decreases the false warnings flagged as a vulnerable dependency. |
| [26] (2021) | Security evaluation of an OSS | The study illustrates a lightweight permission system that safeguards applications from malicious updates from numerous packages. | The research presents a lightweight permission method that sandboxes straightforward third-party packages within the Node.js/npm environment. | Static analysis, sandboxing | Npm/ Node.js | The results show that the proposed framework can safeguard 31.9% of all npm packages, and that 52% of package updates within a one-year span across 120 popular npm packages. |
| [43] (2022) | Vulnerability assessment in imported dependencies | The aim of the study is to support software developers and security specialists in assessing weak link indicators within the npm supply chain. | The work investigates the metadata of 1.63 million JavaScript npm packages, suggesting six indicators of security vulnerabilities in a software supply chain. | Security metrics and metadata analysis | npm/Java Script | The study determined 11 malicious packages, and discovered 2,818 maintainer email addresses linked to expired domains, potentially enabling a malicious actor to hijack 8,494 packages as they might control the npm accounts. |
| [44] (2020) | Vulnerability assessment in used software | This research introduces Mininode, a static analysis method for Node.js applications built to quantify and eliminate superfluous code and dependencies. | The work assessed 672,000 Node.js applications employing Mininode to assess vulnerable dependencies. | Static analysis | Node.js | Mininode identified 1,660 vulnerable packages uploaded from 119,433 dependencies and could remove 2,861 vulnerable dependencies. The proposed tool accurately eliminates 95.4% of packages in the tested Node.js applications. |
| [45] (2019) | Security evaluation of an OSS | This study introduces in-toto, a framework designed to cryptographically provide the SSC's integrity and present software verification from the code implementation phase to the end user. | The in-toto framework was implemented in the context of 30 software supply chain compromises, which had an impact on hundreds of millions of users. | Cryptographic verification | General | The deployment of in-toto shows its proficiency in mitigating software supply chain breaches and maintaining the integrity of software offerings. |

*Table 2 The comparison of studies in SSC security concept*

This section provides an overview of the techniques proposed in academic and grey literature for SSC risk assessment for companies that are software providers (i.e., either software development companies or cloud service providers, see Section 2). Nevertheless, these techniques can also be useful for software users to assess the security of their SSC. The approaches discussed in this section will be used in Section 6 to derive controls that companies can apply to reduce the likelihood of SSC attacks.

In line with established best practices and guidelines (e.g., from MITRE[11] and NIST[12]), the first step of an SSC risk assessment strategy is to produce a software inventory, where all the software installed applications in the company's machines and devices are listed and linked to the organizational operations it is involved in and the organizational assets it accesses.

[11] https://www.mitre.org/our-impact/mitre-labs/systems-engineering-innovation-center/risk-identification

[12] https://csrc.nist.gov/publications/detail/sp/800-30/rev-1/final

The second step is to assess the likelihood that each software will be infected either directly or via an SSC attack. Direct infections can be due to vulnerabilities introduced unintentionally during the software development process. Infections via SSC attacks take place using the strategies described in Section 4 and Section 5. The infection likelihood for a software application depends on its vulnerabilities (see Section 6.1) and the software development practices used within the OSS projects (see Section 6.2) or the third-party software development companies (see Section 6.3) that contributed to the software itself.

The third step consists of estimating the impact of compromised software. Each software listed in the first step is used within one or more organizational operations and has access to a set of organizational assets. A compromised software might (i) stop working, or (ii) behave differently from its specification in terms of functionalities and/or performance, or (iii) be controlled remotely by the attacker, which can also lead to the exfiltration of sensitive information. For each of these three situations, the impact on relevant organizational operations and assets can be analyzed by considering the worst-case scenario.

## 6.1. Vulnerability Assessment

A vulnerability assessment can be carried out on each software installed on the organization's machines and devices. This assessment aims to identify all known vulnerabilities and their severities. Common approaches to finding known vulnerabilities in software include searching on available knowledge bases and relying on software composition analysis (SCA) tools (see Section 6.2.1). Each vulnerability found can be analyzed to assess its severity in terms of exploitability and impact. Note that the impact here refers to the impact the vulnerability would have on the affected software alone, independently from the consequences it would have on organizational operations and assets. Vulnerability severity analysis is discussed in Section 6.2.2.

### 6.1.1. Known Vulnerability Detection

Identifying known vulnerabilities in used software can be achieved using well-established approaches and tools that do not require deep technical knowledge and are well-documented and supported. Several public knowledge bases exist, which collect over time known software vulnerabilities and can be queried to obtain details about specific software versions [32]. The main knowledge base is the U.S. National Vulnerability Database[13] (NVD), managed by NIST, where vulnerabilities are uniquely identified based on the Common Vulnerabilities and Exposure[14] (CVE) system, managed by MITRE. The NVD database offers the possibility to list all the disclosed vulnerabilities for a certain version of a given software application. Further public vulnerability knowledge bases include GitHub security advisories[15] and sonatype OSS index[16], both allowing users to search for vulnerabilities by software application and version [41].

Software composition analysis (SCA) tools produce an inventory of the OSS components included in a software application and report their license and known vulnerabilities. A software development company can use an SCA tool to automate the tracking of the OSS components imported by the software products they develop, as well as ensure they are aware of any known vulnerabilities they are introducing. SCA tools collect details about vulnerabilities both from publicly available knowledge bases and by employing proprietary detection techniques. Imtiaz al. [46] observed experimentally that different SCA tools report different sets of vulnerabilities for the same projects, therefore practitioners

[13] NIST National Vulnerability Database (NVD) https://nvd.nist.gov/

[14] MITRE Common Vulnerability and Exposure (CVE) https://cve.mitre.org/

[15] GitHub security advisories https://github.com/advisories

[16] sonatype OSS index https://ossindex.sonatype.org/

should not rely on a single SCA tool otherwise some vulnerabilities may be missed. The same point was raised by Prana et al [4], as they state that relying solely on public vulnerability databases may cause developers to miss a significant percentage of dependency vulnerabilities. Also, the output of an SCA tool can be used to produce for each product a Software Bill of Materials (SBOM), which gives customers more visibility and transparency on the SSC of the products they use [47].

### 6.1.2. Vulnerability Severity Analysis

Analysing the severity of vulnerabilities helps to prioritize and define a strategy for addressing them. The NVD introduced an approach to assess the severity of a vulnerability, called the Common Vulnerability Scoring System (CVSS)[17], which consists in considering exploitability and impact.

**Exploitability**. The exploitability captures the extent and ease by which a vulnerability can be exploited. The easier a vulnerability can be exploited, the more severe it is considered. The CVSS includes the following metrics to quantify exploitability.

- Attack vector: it expresses how the adversary should reach the vulnerable software to be able to exploit the vulnerability; for example, if the attack can successfully take place remotely over a network, then the exploitability would be higher than if physical proximity were required.
- Attack complexity: it measures the level of difficulty to exploit the vulnerability in terms of the conditions required for the exploitation to succeed; for example, if a certain system configuration is required to exploit a vulnerability, then the attacker should take further action to ensure that that condition holds, therefore the exploitability would be lower.
- Privileges required: it determines whether any particular privilege or permission is needed to exploit the vulnerability; for example, if root access is required to execute an exploit, then the adversary would have to invest further resources to escalate and acquire root privileges, hence the exploitability would be lower.
- User interaction: it indicates whether user participation is needed to exploit the vulnerability; in general, requiring the active involvement of a user calls for the use of social engineering techniques by the attacker, which reduces the chances of success and, thus, makes the exploitability lower.
- Exploit code maturity: it expresses whether a working exploit for that vulnerability exists and is available; for example, if a script implementing a functioning exploit is publicly available and well documented, then the exploitability is higher.

The exploitability of vulnerabilities in dependencies imported from third parties or OSS projects is also determined by whether these dependencies are used in the considered software application. Indeed, if a vulnerable artifact is imported by the application but never used (i.e., no execution path leads to calling any function/method of the vulnerable artifact), then the vulnerability cannot be exploited. In this regard, Pashchenko et al. [33] reported an over-inflation problem in many existing approaches for reporting vulnerable dependencies, because many detected vulnerabilities cannot be exploited in practice. Other academic works [48], [49] proposed approaches to assess the actual exploitability of vulnerabilities by automatically generating test cases aimed at exploiting known vulnerabilities in a dependency; the rationale is, if a vulnerability is exploitable in practice, then its exploitability is higher.

**Impact**. In this context, the impact of a vulnerability represents the consequence of a successful exploit of the software application. As explained at the beginning of Section 6.2, this impact assessment does not take into account consequences to organizational operations or assets, rather it entirely focuses on the effect of the exploitation on the software application only. Although no reviewed paper proposes an approach to assess vulnerability impact, the NVD CVSS introduces three impact assessment metrics to describe the consequences of a successful exploit on the confidentiality, integrity, and availability of the affected software application. Since the NVD provides CVSS scores for almost all known vulnerabilities, software providers can rely on these scores to assess the impact on vulnerable software applications.

[17] Common Vulnerability Scoring System (CVSS) https://nvd.nist.gov/vuln-metrics/cvss

### 6.1.3. Vulnerability management

The assessment of the DevOps strategies employed by a software supplier can be performed using techniques that share similarities with those described in Section 6.3 for the OSS project, with the significant difference that required information would be provided by the supplier itself rather than being collected from public repositories [50]. The only reviewed paper that addresses this point was published by NIST in 2021 [51] and recommends actions a company can take to prevent acquiring malicious or vulnerable software. Based on the definition of these actions, the following indicators can be derived to assess the likelihood that a software supplier could become an entry point for an SSC attack.

1. Whether the supplier enforces a software development lifecycle that integrates security at all stages (i.e., DevSecOps)
2. Whether the supplier includes a vulnerability detection stage within their software development lifecycle
3. Whether identified vulnerabilities are disclosed promptly
4. Whether procedures are in place to respond to identified vulnerabilities, such as patch management policies
5. Whether the supplier has clear procedures and criteria to select their software suppliers and handle vulnerabilities in used dependencies
6. Whether the supplier also releases a component inventory of the software they release, such as the software bill of material (SBOM)

## 6.2. OSS Project Assessment

Analyzing software package suppliers can provide further indicators to estimate the likelihood that they may introduce further vulnerabilities or delay the patching of known vulnerabilities. Software package suppliers can be either OSS projects or third-party software development companies. Given that the large majority of dependencies are OSS packages [30], and that OSS projects provide much more visibility on their internal procedures, many reviewed papers have carried out investigations to identify and measure features that help assess risk due to importing OSS packages. These features aim to capture how well an OSS project deals with fixing vulnerabilities in their dependencies (see Section 6.2.1), whether it exhibits characteristics denoting a higher predisposition to cyber-attacks (see Section 6.2.2), and other aspects such as project importance, whether it is a clone and how much it leverages on third-party code (see Section 6.2.3).

### 6.2.1. Patching practices

Software providers can also assess the risk of using OSS projects by inspecting their patching practices. An OSS project typically depends on several other OSS projects. For example, let us consider an OSS project A that imports another OSS project B as a dependency. When a vulnerability is discovered in B, it takes time for B's developers and maintainers to develop and commit the corresponding patch, release it, publish a security advisory to inform stakeholders (including A), and, finally, further time is required for A’s developers and maintainers to update the dependency itself to include the fix (see Figure 3) [52].

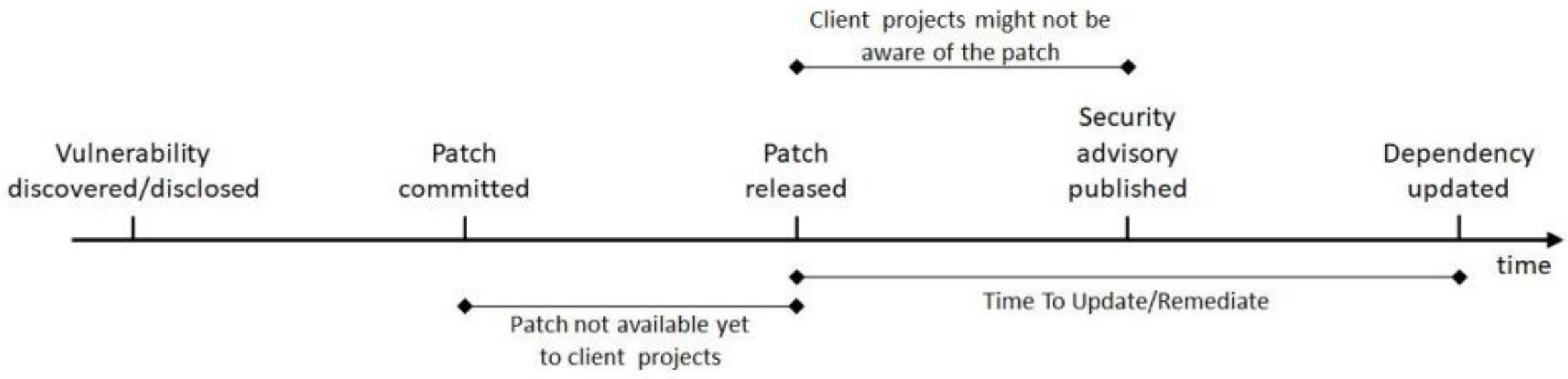

*Figure 6 - Patching timeline for OSS dependencies (adapted from* [12]*)*

The time lag between the fix (i.e., patch committed) and release is an important metric to consider because it represents an additional “window of opportunity” for attackers to exploit the vulnerability [19], [36]. Indeed, committed code in OSS projects can be mined to discover security patches and analyzed to devise the corresponding exploits [53]. During this time, client projects cannot update the vulnerable dependency because the security patch has not been released yet. In their study, Imtiaz et al. [19] found that the median time lag between fix and release is 4 days. Lower values translate to shorter time available for adversaries to exploit the vulnerability, therefore lower likelihood.

The time lag between the patch release and the publication of the corresponding security advisory determines the speed and extent to which client projects are made aware of the security patch and enabled to decide whether to update the dependency [19]. Since the majority of vulnerabilities are fixed silently in OSS projects [37], many software applications take a long time to get patched [54]. Similarly to the previous metric, lower values mean fewer opportunities for attackers to exploit disclosed vulnerabilities [28].

Moreover, once the patch is released, an attacker can be aware of the vulnerable piece of the software, including its location. Consequently, a potential delay in the software updates with new security patches can cause a security attack through the vulnerable part of the software [55]. Avery and Spafford propose a method to mislead the attackers with faux patches composed of additional fake patches along with traditional patches [55]. The authors claim that the faux patches do not reduce the security level of the patches. Instead, it could deceive the attackers and provides extra time for developers to patch their system's vulnerabilities.

A patch release would ideally include information about the security fix implemented and the presence of other unrelated changes if any [42]. These details would allow developers and maintainers of client projects to better understand the advantages (i.e., improved security) and disadvantages (e.g., integrating unrelated changes may require more effort) of updating the corresponding dependency [19]. In particular, patch releases might be bundled together with unrelated breaking changes, which are not backward compatible and would be even more expensive to integrate [56]. These indicators are relevant for risk assessment as (i) better quality of patch release documentation and (ii) avoidance of security releases that also include breaking changes would enable quicker updates to vulnerable dependencies [57].

Mean Time To Update (MTTU) measures the delay between the availability of a patch release and the update of the corresponding dependency. Depending on OSS projects that exhibit lower MTTU reduces the time during which dependencies are left vulnerable. Indeed MTTU was shown to be strongly associated with improved project security [18]. The time to update (TTU) variance is also a valuable metric as it conveys how regularly an OSS project updates its vulnerable dependencies.

Other useful metrics to assess the patching practices of an OSS project include the security history (e.g., how many vulnerabilities it had with a high CVSS score) and the rate of fixes (i.e., how often new releases fix vulnerabilities or bugs) [17].

#### 6.2.2. Weak link signals

Zahan et al. [2] investigated the weak link signals in OSS projects within the npm ecosystem. In terms of SSC security, weak link signals are indicators to estimate the likelihood that an OSS project will be targeted as the entry point of an SSC attack. Although focused on JavaScript projects, the findings of this study are likely to be relevant for OSS projects in general. The authors suggest looking for the following six signals, any of which could identify a project as being an easy target for supply chain attacks.

- Expired Maintainer Domain. If the email address of a maintainer has a domain that is expired, an adversary can buy that domain name from a registrar and take over the maintainer's email account, unless 2-factor authentication is enabled.
- Too many Maintainers. Each maintainer could be targeted by an adversary to obtain privileged access to the project repository, for example by using social engineering techniques. The more maintainers are in a project, the larger the attack surface is and, therefore, the higher the likelihood of a successful SSC attack is.
- Overloaded Maintainer. A person acting as a maintainer for several projects is more susceptible to making errors that could be exploited by an attacker. For example, the attack may infiltrate malware in one of the many dependencies of the maintained projects.
- Too many contributors. Related to the previous signal, maintainers responsible for too many contributors have less time available to properly review received pull requests, which an adversary can exploit to hide malicious code with more chances of success.
- Installation Script. Installation scripts are executed automatically when a dependency is installed. An attacker can target installation scripts to run any malicious commands they want on client projects. Thus, the presence of installation scripts in an OSS project increases both the likelihood and impact of SSC attacks.
- Unmaintained Package. OSS projects that are no longer maintained offer a wider attack surface because (i) attacks targeting those projects are more likely to remain undetected for longer, and (ii) vulnerabilities in those projects are less likely to be fixed promptly. The correlation between SSC attacks and abandoned OSS projects was also explored by Li et al. [32].

#### 6.2.3. Other features

Champion and Hill [58] applied the concept of underproduction (introduced by Gorbati [59]) to measure risk deriving from OSS projects in the SSC. In this context, underproduction occurs when the quality of the project is low compared to its importance, where the quality is measured as the time lag between vulnerability discovery and patch commit (see Figure 3), and importance is evaluated as the number of dependent projects.

Another risk assessment metric is the technical leverage, introduced by Massacci and Pashchenko [60], which is measured as the ratio between third-party code and one's code in an OSS project. Large leverage increases the odds ratio of shipping code with vulnerabilities. A related point was also raised by Gkortzis et al. [31], indeed their study found a strong correlation between a higher number of dependencies and vulnerabilities.

Finally, whether an OSS project is a clone of another project should be checked too [2], as there is a risk that it may have inherited from the original package vulnerabilities that have been fixed there but not in the clone.

## 6.3. Third-party Software Provider Assessment

While the risk due to importing OSS projects can be assessed based on publicly available information, evaluating the exposure to SSC attacks caused by software components or services supplied by third-party software providers requires a slightly different approach [61]. Besides considering how they manage vulnerabilities (see Section 6.1), how they select OSS dependencies (see Section 6.2), and how they select software providers, another key aspect to assess is their cyber posture.

A strong cyber posture reduces the chances that the supplier is successfully targeted by a cyber-attack, which translates to a lower likelihood that malicious code is injected and delivered to downstream customers. A plethora of standards and frameworks have been published that define the controls an organization should have in place to reduce its attack surface and minimize threats, such as the UK Cyber Essentials[18], the NIST Cybersecurity Framework[19], ISO 27001,[20] and ISO 27002[21]. They vary widely in the scope and level of assurance they provide. Detailed comparisons of the most relevant standards and frameworks have been carried out by Taherdoost [62] and Srinivas et al. [63]

[18] NCSC UK Cyber Essentials https://www.ncsc.gov.uk/cyberessentials

[19] NIST Cybersecurity Framework https://www.nist.gov/cyberframework

[20] ISO 27001 https://www.iso.org/isoiec-27001-information-security.html

[21] ISO 27002 https://www.iso.org/standard/75652.html

# 7. Controls to reduce software supply chain attacks likelihood

Building on the risk assessment methodologies introduced in Section 6, suitable controls are identified in this section that companies can use to reduce the risk due to SSC threats. Although a thorough risk management approach should consider both the likelihood and impact of SSC attacks, only the former is analyzed in this section. The *impact* is not investigated here because it depends on the level of exposure of the target to the supplier used as a distribution vector; since this information is specific to each company, extracting representative statistics at the national level is not feasible.

The *likelihood* of different SSC attack techniques was analyzed in Section 5 to determine the top threats to address, based on recent statistics from grey literature. Reducing the likelihood of SSC attacks requires companies to take actions that depend on their role in the SSC. On the one hand, there are several actions that software providers could take to decrease the chances that the products they deliver be used as distribution vectors of a malicious payload. On the other hand, there are also steps software users could take to lessen the probability that they ingest vulnerable or malicious products. Therefore, we propose two sets of controls. The first set of controls outlines steps software providers should take (see Section 7.1), and the second set provides recommended controls for software users to mitigate supply chain risk (see Section 7.2).

## 7.1. Controls for software providers

Software providers can reduce the risk of delivering infected software and services by following the guidelines described in Section 6.3, which cover four areas: cyber posture, vulnerability management, selection of OSS dependencies, and selection of software suppliers. Each of the following sections focuses on one of those areas to derive suitable controls.

### 7.1.1 Controls for cyber posture

Since exploiting cyber and human vulnerabilities in software vendors or cloud providers is one of the most used attack vectors, improving the cyber posture of software providers should be a priority [64]. Two effective controls can be listed as follows:

1. **Cyber Essentials Certification:** a suitable approach to achieve that is by obtaining a *Cyber Essentials* (Plus) certification. Indeed, a recent survey showed that only 6% of UK businesses have Cyber Essentials and 1% of businesses have the Cyber Essentials Plus standard[22]. Cyber Essentials sets out five simple controls to implement: use of firewalls, secure configuration of software and devices, enforcement of user access control policies, installation of anti-malware tools, and management of security updates. Even though the same survey also revealed that 24% of UK businesses have technical controls in place that match all the requirements set out by Cyber Essentials, the numbers are still low; getting more software providers to use Cyber Essentials represents an effective way to improve cyber posture and, consequently, defend better against SSC attacks that exploit cyber vulnerabilities in software providers. Note that the Cyber Essentials scheme also helps against SSC attacks that exploit vulnerabilities in the software development pipeline to bypass authorization or signature verification. Indeed, the management of security updates control aims at fixing these vulnerabilities promptly.
2. **Social Engineering Training:** *training staff* to reduce the chances of success of social engineering techniques against employees of software providers. The UK Cyber Security Breaches Survey [65] reports that only 29% of UK businesses train staff or do mock phishing exercises, which suggests strengthening training is a much-needed countermeasure to put in place. Some of the SSC attacks reviewed relied on phishing to breach a software provider (e.g.,

[22] https://www.gov.uk/government/statistics/cyber-security-breaches-survey-2022/cyber-security-breaches-survey-2022#chapter-4-approaches-to-cyber-security

"*Able Desktop / Operation Stealthy Trident*" and "*Target Supply Chain Attack*"), and they could have been prevented if staff had received more training.

#### 7.1.2 Controls to manage vulnerabilities

Regardless of the particular attack vector used, SSC attacks commonly inject or exploit vulnerabilities in software dependencies. Therefore, a software provider should have controls for vulnerability management integrated within its software development pipeline. Drawing from the approaches introduced in Section 6.1.3, the following controls would be effective to reduce risk due to vulnerabilities in the software developed by the provider.

1. Include a vulnerability detection stage within the software development lifecycle, where established practices can be employed (see Section 6.1.1)
2. Disclose promptly vulnerabilities identified in the software developed
3. Have procedures in place to respond to identified vulnerabilities, such as patch management policies
4. Have procedures in place and criteria set out to select software suppliers and handle vulnerabilities in used dependencies
5. Release a component inventory of the software released, such as the software bill of material (SBOM)

#### 7.1.3 Controls to select OSS dependencies

Since OSS projects are among the top distribution vectors for SSC attacks, a careful selection of the OSS dependencies to include in a software project is of paramount importance [66]. The reviewed literature identifies several aspects to look at to reduce the risk due to OSS dependencies, as discussed in Section 6.2. However, it is challenging to establish controls that can be applied in general to decide whether to include or not a given repository. For example, the metrics described in Section 6.2.1 to assess patching practices, are useful to find correlations with attack likelihood but do not allow setting one-size-fits-all threshold values to use to discard or select OSS projects. The weak link signals in Section 6.2.2 are instead more fitting and the following four allow for the definition of suitable controls that are easy to implement.

1. *Discard projects where the email address of a maintainer has a domain that is expired*. Attacks, where a maintainer's account was compromised, have been successfully launched, such as the "RubyGems Backdoor", and the feasibility of attacks exploiting the expired domain of a maintainer's email address has been shown.
2. *Discard projects that have installation scripts*. This control would have prevented attacks like "VestaCP".
3. *Discard projects that are inactive for more than two years*. This control would have prevented attacks like the one against the popular npm package `ua-parser-js`.
4. *Discard projects that have a maintainer inactive for more than two years*.

#### 7.1.4 Controls to select software suppliers

When a software provider needs to decide whether to pick another company as a supplier of software components or services, the former can check whether the latter implements the controls described in Section 6.2.1

### 7.2. Controls for software users

Software users can decrease their SSC risk by detecting known vulnerabilities in their software estates and fixing them as soon as patches are available to install. This approach is simple to implement and effective against SSC attacks that exploit software vulnerabilities introduced upstream of the supply chain. However, the UK Cyber Security Breaches Survey [65] reports that only 37% of UK businesses have a patching policy in place and apply at least one of the following controls to identify vulnerabilities: undertaking vulnerability audits, performing penetration testing, updating anti-malware software, having a policy that covers damage from using software-as-a-service. Prominent attacks like WannaCry could have been prevented using appropriate known vulnerability detection and patching procedures.

Furthermore, software users can reduce the risk of ingesting vulnerable or malicious software by selecting their software suppliers or cloud providers. According to the UK Cyber Security Breaches Survey [65], only 14% of UK businesses monitor risks from suppliers or the wider supply chain, which suggests there is much room for improvement when it comes to choosing software providers. Besides verifying whether a software provider implements the controls set out in Section 6.2.1, software users can follow the guidelines on supply chain security included in the "10 Steps to Cyber Security" guidance provided by the UK Government.

# 8. Conclusion

This work has provided a comprehensive analysis of the available literature on SSC attacks and risks. For each attack identified, the study has discussed the corresponding attackers' entry points within the SSC. In addition, a review of SSC risk assessment methodologies has been provided. Our results show that the most common entry points for SSC attacks are OSS projects, software vendors, and cloud providers. It has also been noted that there is a significant increase in the rate of SSC attacks with OSS projects as entry points. This finding is especially concerning given the fact that, nowadays, a large portion of software development projects rely heavily on the use of OSS components. This research has focused on these more critical attack and distribution vectors to propose a set of control measures, which include mitigation techniques for both software providers and users. For future work, we plan to develop an automated tool for self-assessment of SSC attacks to help developers better protect their software products.

# Acknowledgment

This work is supported by DCMS, Department for Digital, Culture, Media and Sport.